\documentclass[final,sort&compress]{aipproc}
\layoutstyle{8x11single}
\usepackage{epstopdf}

\begin{document}
\bibliographystyle{MoH2014}
\title{The avian tectorial membrane: Why is it tapered?}
\classification{}
\keywords{}

\author{Kuni H Iwasa}{
   address={Otolaryngology \& Head and Neck Surgery, Stanford University, Stanford, CA 93405, USA}, altaddress={NIDCD, NIH, Bethesda, MD 20892, USA}
}

\author{Anthony J Ricci}{address={Otolaryngology \& Head and Neck Surgery, Stanford University, Stanford, CA 93405, USA},
altaddress={Molecular \& Cellular Physiology, Stanford University, Stanford, CA 93405, USA}}

\begin{abstract}
While the mammalian- and the avian inner ears have well defined tonotopic organizations as well as hair cells specialized for motile and sensing roles, the structural organization of the avian ear is different from its mammalian cochlear counterpart. Presumably this difference stems from the difference in the way motile hair cells function. Short hair cells, whose role is considered analogous to mammalian outer hair cells, presumably depends on their hair bundles, and not motility of their cell body, in providing the motile elements of the cochlear amplifier. This report focuses on the role of the avian tectorial membrane, specifically by addressing the question, ``Why is the avian tectorial membrane tapered from the neural to the abneural direction?'' 
\end{abstract}

\maketitle
 
 \section{Introduction}
The avian and mammalian ears have similarities and differences. The innervation pattern of avian tall hair cells resembles that of mammalian inner hair cells in having primarily afferent innervation, suggesting their role as the primary sensor. In contrast, the innervation pattern of short hair cells in the avian ear is exclusively efferent and resembles that of outer hair cells, suggesting the role of motor element in the cochlear amplifier \cite{Hirokawa1978,Gleich1989,Fischer1992}. Since bending of hair bundles (including tilting motion of the apical surface of hair cells \cite{Beurg2013}) is the only motile response in short hair cells, such a physiological role requires that the hair bundles of short hair cells exert force between the tectorial membrane and the basilar papilla. 

There are marked differences. One such difference is the absence of pillar cells and the tunnel of Corti in the avian ear. In addition, the presence of electrical tuning \cite{Fuchs1988} and absence of somatic motility in avian hair cells make them apart. Yet another is the morphology of the tectorial membrane (TM). Is it true that the difference in the motile element of the cochlear amplifier is the reason for the structural difference?  Here this issue is examined by focusing upon the morphology of the tectorial membrane.

%\begin{wrapfigure}{r}{0.48\textwidth}
\begin{figure}[tb]
%  \begin{center}
  \includegraphics[width=0.42\textwidth]{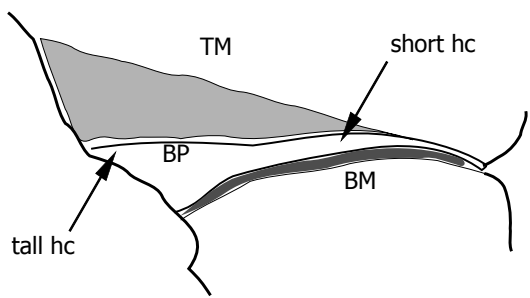}
%\includegraphics[width=0.42\textwidth]{figs/avian-ear.pdf}
%  \end{center}
  \caption{\small{The anatomy of the avian inner ear. TM: the tectorial membrane, BM: the basilar membrane, BP: the basilar papilla. After O.\ Gleich and G.\ A.\ Manley \cite{gle-man2000}.}}
    \label{fig:anatomy}
\end{figure}    
%\end{wrapfigure}
\indent The stiffness of the avian TM would be low compared with the mammalian one, being devoid of collagen  \cite{Pickles1990}. It shows perforations reminiscent of Swiss cheese in electron micrographs due to fixation instead of continuous low density mass as observed in unfixed preparations \cite{Runhaar1989a}. Since collagen is largely responsible for the mechanical anisotropy of the mammalian TM \cite{Masaki2009}, its absence would make the material properties of the avian TM very different from the mammalian one, in which stiffness gradient and the significance of the resonance frequency has been documented \cite{Richter2007,Ghaffari2007,Jones2013,Gavara2012,Lamb2011}. 

The avian TM is tapered toward the abneural side (Fig. \ref{fig:anatomy}). The steepness of tapering increases toward the basal end with higher best frequencies  \cite{Koeppl1998}. In contrast to the flat and featureless surface of the mammalian tectorial membrane \cite{lim1986}, the avian TM has an elaborate structure on the side facing the basilar papilla. A thin veil-like structure descends to the basilar papilla, surrounding each hair cell \cite{Goodyear1994}. A dome like recess above each hair cell and each hair bundle makes contact with the tectorial membrane at the dome \cite{Goodyear1994}. What is the functional significance of these structural features? 

Like outer hair cells in the mammalian ear, short hair cells are located nearly above the middle of the basilar membrane, the movement of which in the up-down direction \cite{Gummer1987} would effectively transmit bending stress to the short hair cells' hair bundles. In contrast, tall hair cells, which functions as the sensor, are located away from the center of the basilar membrane, at a location seemingly unsuitable to  sense the motion of the basilar membrane directly. Thus it is likely short hair cells are involved in amplification by a hair bundle active process \cite{cho-hud1998,Tinevez2007,Sul2009c}, transmitting the mechanical vibration of the basilar membrane to the hair bundles of tall hair cells. The problem is how that can happen? To address this question, it requires to identify the vibrational mode, which involves the basilar membrane, the basilar papilla, and the TM. Since reported measurements are limited to the basilar membrane with M\"ossbauer technique \cite{Gummer1987}, we need to assume how other elements could move. Since not much else is known for certain, the models considered herein are largely based on morphological observations and rather speculative.

Since hair cell bundles are stimulated by shear, it would be logical to assume that shear motion between the basilar papilla and the TM stimulates the hair bundles of tall hair cells, similar to the mammalian ear. However, there are clear differences. The orientation of hair bundles in the avian ear is not as uniform as in mammalian cochlea. Although the orientation is relatively uniform in the basal region,  it forms domain structure in a more apical region \cite{Koeppl1998,gle-man2000}. In the following, we examine the mode of motion that is consistent with the bundle alignment in the basal area, then consider a possible mechanism for domain formation in more apical region.

\section{Sliding mode of motion}

%\begin{wrapfigure}{l}{0.3\textwidth}
\begin{figure}[h]
%  \begin{center}
  \includegraphics[width=0.22\textwidth]{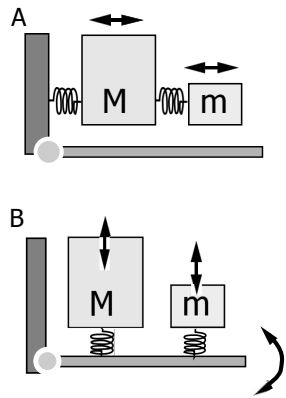}
%\includegraphics[width=0.42\textwidth]{figs/modes.pdf}
%  \end{center}
  \caption{\small{Two possible modes of motion of the avian inner ear. Two squares represent two parts of the tectorial membrane. The smaller one on top of short hair cells and the larger one on top of tall hair cells.}}
    \label{fig:modes}
\end{figure}
%\end{wrapfigure}

Electron micrographs show that hair bundles are oriented primarily in the direction orthogonal to the longitudinal axis of the basilar papilla, particularly in the basal higher frequency region. This orientation is consistent with the shear between the tectorial membrane and the basilar papilla, analogous to the mammalian ear (Fig.\ \ref{fig:modes} A).

However, there are appreciable differences between the two. The avian ear does not have pillar cells or a tunnel of Corti. Thus it may not have well-defined pivoting points for creating shear from the up-down motion of the basilar membrane. In addition, the tectorial membrane (TM) is tapered, making it thicker on top of tall hair cells. The TM is attached to a wall, where it is thickest, nearer to tall hair cells. How can such a morphology be consistent with the assumption that short hair cells serve as the cochlear amplifier? To explain this puzzle, Charles Steele suggested that the TM should have resonance near the best frequency \cite{Steele1997}. Indeed, resonance would make it work. However, a question still remains. Why should it tapered toward short hair cells? The avian TM works despite its shape? Is there an advantage of this morphology? 

Since so many factors remain uncharacterized, it would be advantageous to examine the simplest possible model to obtain a conceptual understanding of the issue rather than more complex ones, which would require many more details. For this reason, two-mass models are used here for describing the TM (Fig.\ \ref{fig:modes}). One has a smaller mass $m$, located at the top of short hair cells and another with a larger mass $M$, located at the top of tall hair cells.

The small mass $m$ is connected to the larger mass $M$ with a spring with stiffness $k$ and the larger mass $M$ is connected with the wall by another spring with stiffness $K$. Let $X$ the displacement of the bigger mass from its equilibrium position, and that of smaller one be $x$. Let $\eta_2$ be viscous drag on mass $M$ and $\eta_1$ be negative drag on mass $m$ to mimic the amplifying role of short hair cells (Fig.\ \ref{fig:modes} A). The set of equations is given by,
\begin{eqnarray}
\left(m\frac{d^2}{dt^2}-\eta_1\frac{d}{dt}\right)x-k(X-x)=f_0\exp[i\omega t] \\
\left(M\frac{d^2}{dt^2}+\eta_2\frac{d}{dt}+K\right)X-k(X-x)=0,
\label{eq:motion}
\end{eqnarray}
where $f_0\exp[i\omega t]$ is a periodic external force of angular frequency $\omega$ applied to mass $m$, which is located just above the center of the basilar membrane. Here it is assumed that up-down motion of the basilar membrane results in this external force. Following Steele \cite{Steele1997}, let $\omega_0$ be the resonance frequency of the tectorial membrane, i.e. $\omega_0^2=K/M=k/m$ and seek a steady state solution. 

%\begin{wrapfigure}{r}{0.45\textwidth}
\begin{figure}
%  \begin{center}
 \includegraphics[width=0.4\textwidth]{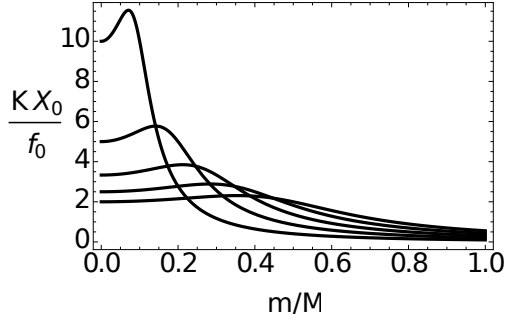}
%\includegraphics[width=0.45\textwidth]{figs/drag-dependence2.pdf}
%  \end{center}
  \caption{\small{Force amplitude $K X_0/f_0$, normalized to that of the driving external force, applied to the mass $M$ at $\omega=\omega_0$, is plotted against the mass ratio $m/M$ using the reduced drag coefficient $\overline{\eta}=\eta/(M\omega_0)$ as a parameter, which ranges from 0.1 to 0.5 incremented by 0.1. The peak is at small values of  mass ratio $m/M$ for the small value of $\overline{\eta}$ and systematically shifts to larger values as $\overline{\eta}$ increases.}}
    \label{fig:drag}
\end{figure}
%\end{wrapfigure}

Under this assumption, the force applied to the mass $M$ is plotted for a special case of $\eta_1=\eta_2=\eta$ (Fig.\ \ref{fig:drag}). For a given value of $M$, this force is larger for smaller values of the reduced drag $\overline{\eta}$ where the ratio $m/M$ is smaller. Since the reduced drag is defined by $\overline{\eta}=\eta/(M\omega_0)$, this parameter is smaller for higher frequency $\omega_0$. Thus the plot shows that the force is maximal for smaller ratio $m/M$ at larger frequencies. This result is consistent with the experimental finding that the sharpness of tapering of the TM increases from the apical to the basal end \cite{Smith1985,Koeppl1998}. One can argue that the force relevant to the transduction of tall hair cells would be better related to drag $\eta dX/dt$ rather than the force $KX_0$. That does not change the optimal mass ratio because drag is represented by $\overline{\eta}KX_0$, maximizing at the same mass ratio. 
The condition for resonance is not very strict because those peaks exist for $1 \leq \omega/\omega_0 \leq 1.4$.

%%%%%%
\section{Jumping mode of motion}

\textbf{Domain structure}

While hair bundles in the basal area of the basilar papilla are aligned perpendicular to the longitudinal axis of the papilla, those in more apical area are aligned in a patchwork fashion, particularly in the middle between neural and abneural ends. How can this pattern be explained?

Suppose the motion of the tectorial membrane relative to the basilar papilla is perpendicular to the surface of the basilar papilla (Fig.\ \ref{fig:modes} B), how do hair bundles bend?  Each hair bundles is attached to the tectorial membrane at the edge of a dome-like recess \cite{Goodyear1994}. The fibrous veils, which descend from the tectorial membrane and reach the basilar papilla, surround each hair cell at its margins. Some parts of the seemingly extendable veils form triangles with hair bundles and parts of the basilar papilla \cite{Goodyear1994}, which is much denser and therefore expected to be much stiffer. This structure could couple elastic elongation and shortening of the veils, resulting in a bending of the hair bundles. 

Since bending of a hair bundle results in local displacement of the TM, bending of the neighboring hair bundles in the same direction is energetically favorable if those hair bundles are oriented in the same direction. For this reason, up-down motion of the TM favors domain formation in hair bundle orientations. The size of domains is limited to even out those local deformations of TM to make the average motion in the up-down direction.\\

\noindent \textbf{Equations of motion}

In this mode of motion, we assume for simplicity, that two masses do not interact directly, but only indirectly being mediated by the basilar papilla, which pivots at the supporting point (Fig.\ \ref{fig:modes} B). Let $X_0$ the distance of the large mass $M$ from the pivoting point, and $x_0$ that of the smaller mass $m$. Notice that $x_0>X_0$ and for that reason, the torque is more effectively produced by the force applied by the mass $m$. The equations of motion would be given by,
\begin{eqnarray} \label{eq:ud1}
\left(M\frac{d^2}{dt^2}+\eta_1\frac{d}{dt}+K_p\right)Y-K_p X_0\theta&=&0,\\  \label{eq:ud2}
\left(m\frac{d^2}{dt^2}+\eta_2\frac{d}{dt} +k_p\right)y-k_p x_0\theta&=&0, \\  \label{eq:ud3}
\left(I\frac{d^2}{dt^2}+\eta_3\frac{d}{dt} +k_p x_0+K_p X_0+\kappa\right)\theta-K_p Y-k_p y&=&f_0\exp[i\omega t],
\end{eqnarray}
where $I$ is the moment of inertia of the basilar papilla, $\theta$ the angle of rotation of the basilar papilla around its pivoting point and $\eta_j$'s (j=1,2,3) are either positive or negative, depending on whether they have damping or amplifying effect. The quantity $y$ is the perpendicular displacement of the mass $m$, and $Y$ that of the mass $M$. Here $k_p$ and $K_p$ represent elasticity due to hair bundles and veils. The quantity $\kappa$ represents the elastic elements not associated with hair bundles.

\begin{figure}
%  \begin{center}
 \includegraphics[width=0.4\textwidth]{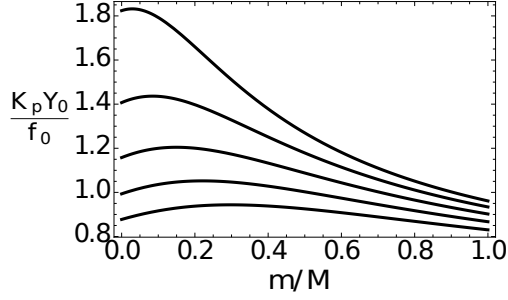}
%  \end{center}
  \caption{\small{Force amplitude $K_p Y_0/f_0$, normalized to that of the driving external force, is plotted against the mass ratio $m/M$, assuming $\eta_1=-\eta_2=\eta$, $K_p=k_p$, $\omega_0^2=K_p/M$, and $\omega/\omega_0=1$. The reduced drag coefficient $\overline{\eta}=\eta/(M\omega_0)$ is a parameter, which ranges from 0.3 to 0.7 incremented by 0.1. The peak is at small values of  mass ratio $m/M$ for small values of $\overline{\eta}$. The peak systematically shifts to larger values as $\overline{\eta}$ increases. $X_0/(X_0+x_0)=0.1$}}
    \label{fig:jump}
\end{figure}

Since this is a three body problem, which may not have a periodic solution in general, we examine a special case. Since we are interesting in the mass ratio $m/M$, we regard the basilar papilla as mediating energy transfer between the smaller mass $m$ and the bigger mass $M$, letting $I=\eta_3=\kappa=0$. The longer lever arm of the hair bundles of short hair cells have is advantageous for short hair cells to function as an amplifier. Under this condition, the force applied to the mass $M$ potted against the mass ratio $m/M$ has a maximum for a given $\overline{\eta}$, similar to the sliding mode (Fig.\ \ref{fig:jump}).

%\begin{wrapfigure}{l}{0.45\textwidth}
%\end{wrapfigure}

We assumed that short hair cells provide negative damping whereas tall hair cells do not.  This requires that active force generation at short hair cells is greater than in tall hair cells. In addition, the source of viscous drag is important. Up-down motion of the tectorial membrane with respect to the basilar papilla requires volume changes in the gap between the tectorial membrane and the basilar papilla, even though changes are small. Since the soft and possibly porous structure of the tectorial membrane may not allow fluid movement at auditory frequencies, the fluid in the gap is likely go through veils surrounds each hair cells, creating an extra viscous drag. In general, such drag would reduce the energy efficiency. However, it could be beneficial in this particular situation if it makes the drag of tall hair cells net positive and that of short hair cell net negative.

\subsection{Dominance of modes}
The modes of motion described above are not mutually exclusive and are likely coexist. The observation that basal hair bundles are well aligned than more apical bundles and that basal hair bundles have flatter appearance than more apical ones \cite{Koeppl1998,gle-man2000} could indicate a relative significance of the sliding mode at higher frequencies. It is likely that in more apical area the jumping mode, which is consistent with domain formation, is dominant. The tapering of the TM is a feature optimizing the sensitivity of tall hair cells in the presence of active motion of the hair bundles of short hair cells.

%\hrulefill

\section{Discussion}
One could argue that studying the modes of motion in the avian inner ear could be premature: First of all, very few physiological data are available. The only data on the vibration of the basilar membrane was published in 1987 by Gummer at al.\ using the M\"ossbauer effect \cite{Gummer1987}, showing traveling wave and that the amplitude is somewhat smaller compared with the mammalian counterpart. The mechanical properties of the basilar papilla nor of the TM have not been measured. Moreover, studying the modes of motion may not be easy. Even in the mammalian cochlea, which has been studied much more extensively, the details are becoming clearer only recently.

Nonetheless the avian ear would be an interesting subject for a couple of reasons. One such reason is that the avian basilar papilla has some resemblance to the mammalian ear, such as tonotopic organization and differentiated roles of two types of hair cells, and yet it has much simpler morphology, such as the absence of pillar cells and the tunnel of Corti besides different TM morphologies, possibly suggesting simpler modes of motion. In addition, its cochlear amplifier must depend solely on an active process in hair bundles, which may be based on fast adaptation \cite{cho-hud1998,Tinevez2007,Sul2009c} and possibly, in part, avian prestin \cite{Beurg2013}. Also the electrical properties of chick hair cells \cite{Fuchs1988}
%,Samaranayake2004
are quite different from mammalian hair cells in having an intrinsic electrical tuning mechanism that likely in part shapes the mechanical components, whether they are the hair bundle or prestin \cite{Beurg2013}. For this reason, the mechanical motion in the avian ear does not have to be as sharply tuned as in the mammalian one. It is possible that the relatively simple structure of the avian ear could possibly be advantageous for identifying the modes of vibration.

Assuming a local TM resonance \cite{Steele1997} is attractive in view of its role in the mammalian cochlea \cite{Richter2007,Ghaffari2007,Jones2013,Lamb2011}. For the jumping mode of motion, it would be reasonable because hair bundles, which have systematic gradient \cite{Koeppl1998,gle-man2000}, would make a significant contribution to the stiffness. How realistic is the assumption of resonance for the sliding mode of motion? Could Young's modulus, density, and the dimension of the TM be compatible with the assumption of resonance from the basal end to the apical end? Can a gradient of the mechanical properties of the TM exist? These issues are perhaps experimentally testable.

The difference between short and tall hair cells is another requirement for our models models presented here. Specifically, our models require that the net effect of tall hair cells should be drag and that of short hair cells should be negative drag. This requires substantial difference in the magnitude of active force production in the two hair cell types. Indeed, some difference in bending responses, both voltage-induced and not, in these hair cell types has been observed \cite{Beurg2013}. However, it remains unclear whether or not the differences are large enough to satisfy our requirement.

There are other possible modes of motion that are not considered here. The degree of simplification of the present models in describing the TM is obvious. In addition, there are other factors to consider. For example, does the basilar membrane move together with the basilar papilla as a solid body or can they possibly have a difference in the phase as well as amplitude? Photomicrographs show an appreciable space between the basilar papilla with embedded hair cells and the basilar membrane \cite{Koeppl1998}. In addition, these two largely parallel structures appear to have separate anchoring points on the neural side. This feature could possibly allow an external force to drive basilar papilla more than the tectorial membrane. 

\begin{theacknowledgments}
Thanks to Drs.\ Christine K\"oppl and Goeff Manley of Oldenburg, Germany and Charles Steele of Stanford for discussion.
\end{theacknowledgments}

%\bibliography{/Users/kuni/ms/bib/cilia,/Users/kuni/ms/bib/ohc}  

\end{document}